\documentclass[twocolumn]{aastex631}

\newcommand\jykmsmath{\rm{[Jy\ km\ s}^{-1}\rm{]}}

\newcommand\kms{\rm{km\ s}^{-1}}

\submitjournal{ApJ}

\shorttitle{ALMA 300 pc resolution imaging of a z=6.79 quasar}
\shortauthors{Meyer et al.}

\graphicspath{{./}{figures/}}

\begin{document}

\title{ALMA 300 pc resolution imaging of a z=6.79 quasar: no evidence for supermassive black hole influence on the [\ion{C}{2}] kinematics}

\correspondingauthor{Romain A. Meyer}
\email{romain.meyer@unige.ch}

\author[0000-0001-5492-4522]{Romain A. Meyer}
\affiliation{Max Planck Institut f\"ur Astronomie, K\"onigstuhl 17, D-69117, Heidelberg, Germany}
\affiliation{Department of Astronomy, University of Geneva, Chemin Pegasi 51, 1290 Versoix, Switzerland}
\author[0000-0002-9838-8191]{Marcel Neeleman}
\affiliation{National Radio Astronomy Observatory, 520 Edgemont Road, Charlottesville, VA, 22903, USA}
\affiliation{Max Planck Institut f\"ur Astronomie, K\"onigstuhl 17, D-69117, Heidelberg, Germany}
\author[0000-0003-4793-7880]{Fabian Walter}
\affiliation{Max Planck Institut f\"ur Astronomie, K\"onigstuhl 17, D-69117, Heidelberg, Germany}
\author[0000-0001-9024-8322]{Bram Venemans}
\affiliation{Leiden Observatory, Leiden University, PO Box 9513, 2300 RA Leiden, The Netherlands}

\begin{abstract}
We present Atacama Large Millimeter/submillimeter Array (ALMA) [\ion{C}{2}] 158 $\mu \rm{m}$ and dust continuum observations of the $z=6.79$ quasar J0109--3047 at a resolution of $0\farcs045$ ($\sim$300 pc). The dust and [\ion{C}{2}] emission are enclosed within a $\sim 500\, \rm{pc}$ radius, with the central beam ($r<144\ \rm{pc}$) accounting for $\sim$25\%  (8\%) of the total continuum ([\ion{C}{2}]) emission. The far--infrared luminosity (FIR) density increases radially from $\sim$5 $\times 10^{11} L_\odot\ \rm{kpc}^{-2}$ to a central value of $\sim$70 $\times 10^{11} L_\odot\ \rm{kpc}^{-2}$ (SFRD $\sim$50-700 $M_\odot\ \rm{yr}^{-1}\ \rm{kpc}^{-2}$). The [\ion{C}{2}] kinematics are dispersion-dominated with a constant velocity dispersion of $137 \pm 6 \,\kms$. The constant dispersion implies that the underlying mass distribution is not centrally peaked, consistent with the expectations of a flat gas mass profile. The lack of an upturn in velocity dispersion within the central beam is inconsistent with a black hole mass greater than $M_{\rm{BH}}<6.5\times 10^{8}\ M_\odot\ (2\sigma$ level), unless highly fine-tuned changes in the ISM properties  conspire to produce a decrease of the gas mass in the central beam comparable to the black hole mass. Our observations therefore imply either that a) the black hole is less massive than previously measured or b) the central peak of the far-infrared and [\ion{C}{2}] emission are not tracing the location of the black hole, as suggested by the tentative offset between the near-infrared position of the quasar and the ALMA continuum emission.
\end{abstract}


\section{Introduction} \label{sec:intro}

Since their discovery in the early 2000s, luminous quasars at $z>6$ \citep[][]{Fan2006} have provided new challenges and insights into the first phases of supermassive black hole (SMBH) formation and growth. Indeed, luminous $z>6$ quasars are found to be powered by Eddington--accreting SMBHs with $M_{\rm{BH}} \sim 10^{9} M_\odot$, which imply either very large black hole seeds ($10^{3}-10^{5} M_\odot$) at $z>15$ and/or earlier super--Eddington accretion episodes \citep[][]{DeRosa2014, Mazzucchelli2017, Yang2021b,Farina2022}. Although both massive seeds or super--Eddington accretion scenarios are seen in simulations invoking various physical mechanisms \citep[see, e.g., ][for comprehensive reviews]{Haiman2004,Overzier2009,Volonteri2010,Latif2016,Inayoshi2020, Volonteri2021}, observational evidence for one or either channel is currently lacking. The recent discoveries of $z>7$ quasars with $M_{\rm{BH}} \sim 10^{9} M_\odot$ have only accentuated this issue \citep[][]{Mortlock2011, Banados2018, Yang2018a, Yang2020, Wang2018a,Wang2020, Wang2021}. Additionally, kinematical studies of the [\ion{C}{2}] line of high--redshift quasar host galaxies have shown that their SMBH are typically overmassive compared to the local galaxy--BH mass relation \citep[e.g.,][]{Pensabene2020,Neeleman2021}.

An alternative to this “early SMBH growth/seed problem" is the possibility that their SMBH masses are overestimated. Indeed, black hole masses at $z>6$ are currently measured in most cases via the single--epoch virial estimator using the width of the rest-frame \ion{Mg}{2} broad emission line \citep[e.g.,][]{Vestergaard2009,Shen2011}. The \ion{C}{4} line is also possibly suitable for this measurement \citep[e.g.,][]{Vestergaard2006}, but the larger scatter of the scaling relation and its dependence on the blueshift of the line make it a less reliable estimator \citep[][]{Coatman2017, Mejia-Restrepo2018}. This is reinforced by several studies suggesting that the median blueshift of \ion{C}{4} is evolving in $z>6$ quasars \citep[][]{Mazzucchelli2017,Meyer2019b, Schindler2020, Yang2021b}. It remains to be seen whether more direct measurements from the $\rm{H}\alpha$ or $\rm{H}\beta$ lines with \textit{JWST} will agree with the current rest-frame UV estimates \citep[see e.g.][]{Yang2023}. Nonetheless, it is possible that low--redshift empirical relations might not apply to the most luminous quasars in the first billion years of the Universe. Given the implication for early black hole growth and galaxy--SMBH co--evolution, an independent measurement of the black hole masses in $z>6$ quasars is thus highly desired. 

One of the most accurate methods to measure the mass of a black hole is to detect the kinematical signature of material directly influenced by the gravity of the black hole \citep[e.g.,][]{Dressler1988}. This region is called the black hole's sphere of influence and for a 10$^9\ M_\odot$ black hole corresponds to a physical size of about $200-400\ \rm{pc}$ (in the absence of a significant gas/stellar mass component). Stellar kinematics have long been the gold standard for accurately measuring black hole masses in our Galaxy \citep[e.g.,][]{Ghez2008,Gillessen2009} and nearby galaxies \citep[e.g.,][]{Magorrian1998, Verolme2002, Gebhardt2003}. However, recent work have shown the feasibility of far--infrared (FIR) molecular gas tracers for making this measurement \citep[e.g.,][]{Davis2013, Davis2017}. For luminous quasars at $z>6$, where the stellar emission is outshone by the accreting SMBH in the UV, only FIR atomic and molecular lines can hope to probe the black hole's sphere of influence. However, the resolution and sensitivity that is required to make these observations remains challenging, and therefore only few such observations of $z>6$ quasars (at $\lesssim 300\, \rm{pc}$ resolution) exist \citep[][]{Venemans2019,Walter2022}.

These pioneering observations have shown that reaching sub--kpc resolution is not always sufficient to constrain the kinematics of the black hole's sphere of influence in order to estimate its mass. For example, \citet[][]{Venemans2019} presented $400\ \rm{pc}$ resolution [\ion{C}{2}] imaging of the $z=6.6$ quasar J0305--3150. These observations could not constrain the black hole mass due to the highly perturbed kinematics of the merging system studied and the high gas mass in the center. Recent $200\ \rm{pc}$ resolution observations of the $z=6.9$ quasar J2348--3054 by \citet[][]{Walter2022} could also not resolve the sphere of influence. This was due to the unexpectedly high mass concentration of dust and gas in the inner hundreds of parsecs of the quasar host galaxy, which far outweigh the mass of the SMBH. It is thus worth noting that finding a suitable system is as important as reaching the high--resolution and sensitivity necessary for a kinematical black hole mass measurement.

In this paper, we present new observations of the [\ion{C}{2}] and dust continuum emission of the $z=6.79$ quasar, J0109--3047, at a resolution of $0\farcs045$ ($\sim 300\, \rm{pc}$ at $z=6.79$). Our goals are to study the dust distribution and ISM kinematics, and to resolve the black hole sphere of influence. We summarize the observations and data reduction in Section \ref{sec:data} and describe the compact nature of the source in Section \ref{sec:ISM_results}, including SFR and FIR surface densities estimates. We discuss the [\ion{C}{2}] kinematics and the implied constraints on the SMBH mass in Section \ref{sec:soi}. We speculate on the possibility that the quasar is offset from the central host galaxy in \ref{sec:offset}  before concluding with a short summary in Section \ref{sec:conclusions}. Throughout this paper we use a concordance cosmology with $H_0=70\  \kms \rm{Mpc}^{-1}, \Omega_M = 0.3, \Omega_\Lambda=0.7$. Accordingly, $1"$ at $z=6.79$ corresponds to $5.32$ proper kpc. 

\section{Observations and data reduction} \label{sec:data}

We observed the $z=6.79$ quasar J0109--3047 with the Atacama Large Millimeter/submillimeter Array (ALMA) in configuration C43-8 between 2021 October 14 and 25. The observations were carried out for a total of $7.3h$ on--source in Band 6 and targeted the redshifted [\ion{C}{2}] emission line and the underlying continuum at $\nu_{\rm{rest}} \sim 244\ \rm{GHz}$. J0006--0623 was used for bandpass and amplitude calibration and J0106--2718 for phase calibration. The data presented in this work were reduced using standard CASA (6.2.1.7) routines and combined with previous low--resolution observations from cycle 1,2,3 \citep[with beam sizes $0\farcs6 - 0\farcs2$,][]{Venemans2016, Venemans2020}. Imaging was performed using Briggs weighting with robust parameter $r=0.5$. The resulting synthesized beam has a major axis $a=0\farcs049$ and minor axis $b=0\farcs041$, leading to a beam area of $\Omega=0.0023\ \rm{arcsec}^{2}$. This corresponds to an effective radius of $0\farcs027$, or $144\ \rm{pc}$ at $z=6.79$. 
\begin{figure*}
    \centering
    \includegraphics[width=0.9\textwidth]{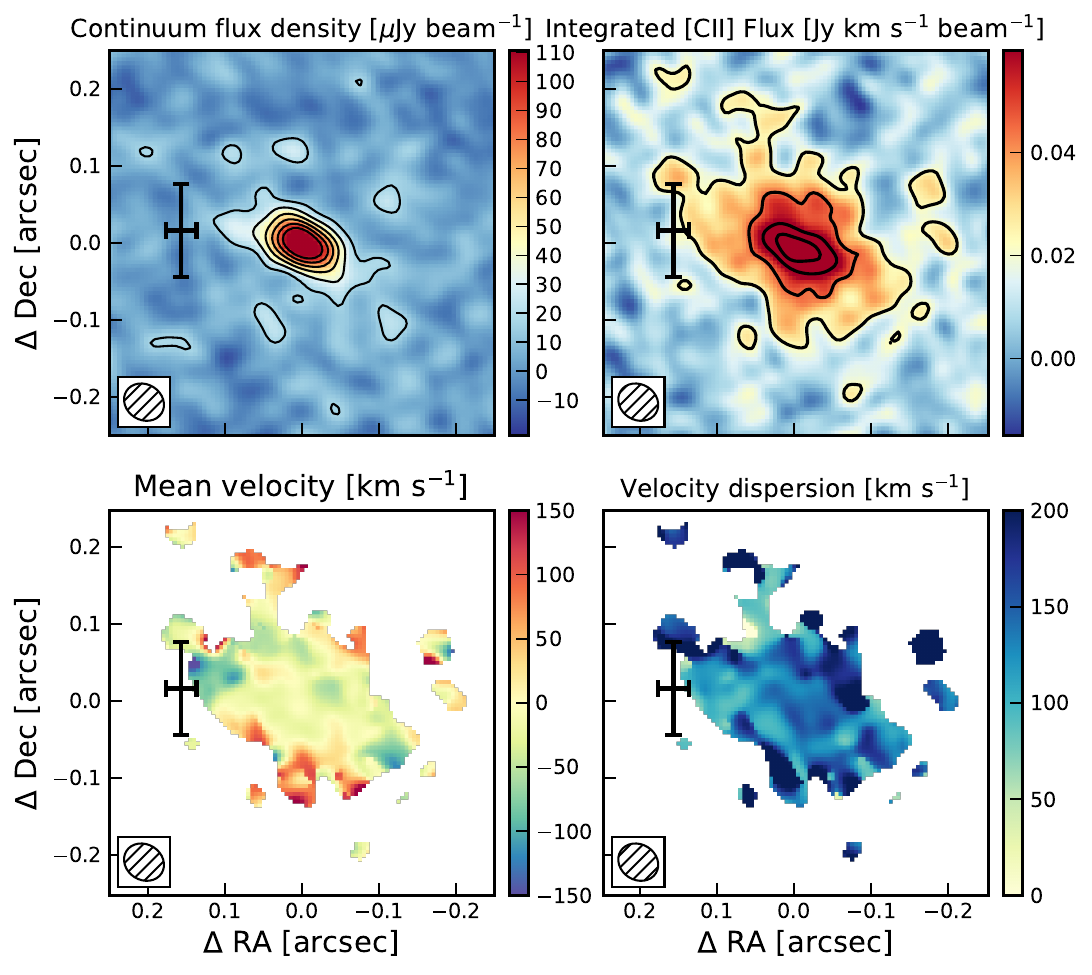}
    \caption{\textbf{Top row:} Continuum emission and velocity--integrated [\ion{C}{2}] emission maps of J0109--3047 at $\sim 300$ pc resolution. The black crosses indicates the NIR position of J0109--3047 based on the GAIA--corrected positions of stars close to the quasar in the near-infrared images \citep[][]{Venemans2020}. \textbf{Bottom row: } [\ion{C}{2}] mean velocity and velocity dispersion fields.  All maps are produced from the complete Cycle 1,2,3 and 8 data. The synthesized beam is $0\farcs049\times0\farcs041$. The continuum and [\ion{C}{2}] velocity--integrated emission maps are produced using residual scaling. The contours are given at $(3,6,9,12)\sigma$ as black lines (no negative $(-6,-3)\sigma$ emission is detected).  \label{fig:maps} }
\end{figure*}
\begin{figure*}
    \centering
    \includegraphics[width=\textwidth]{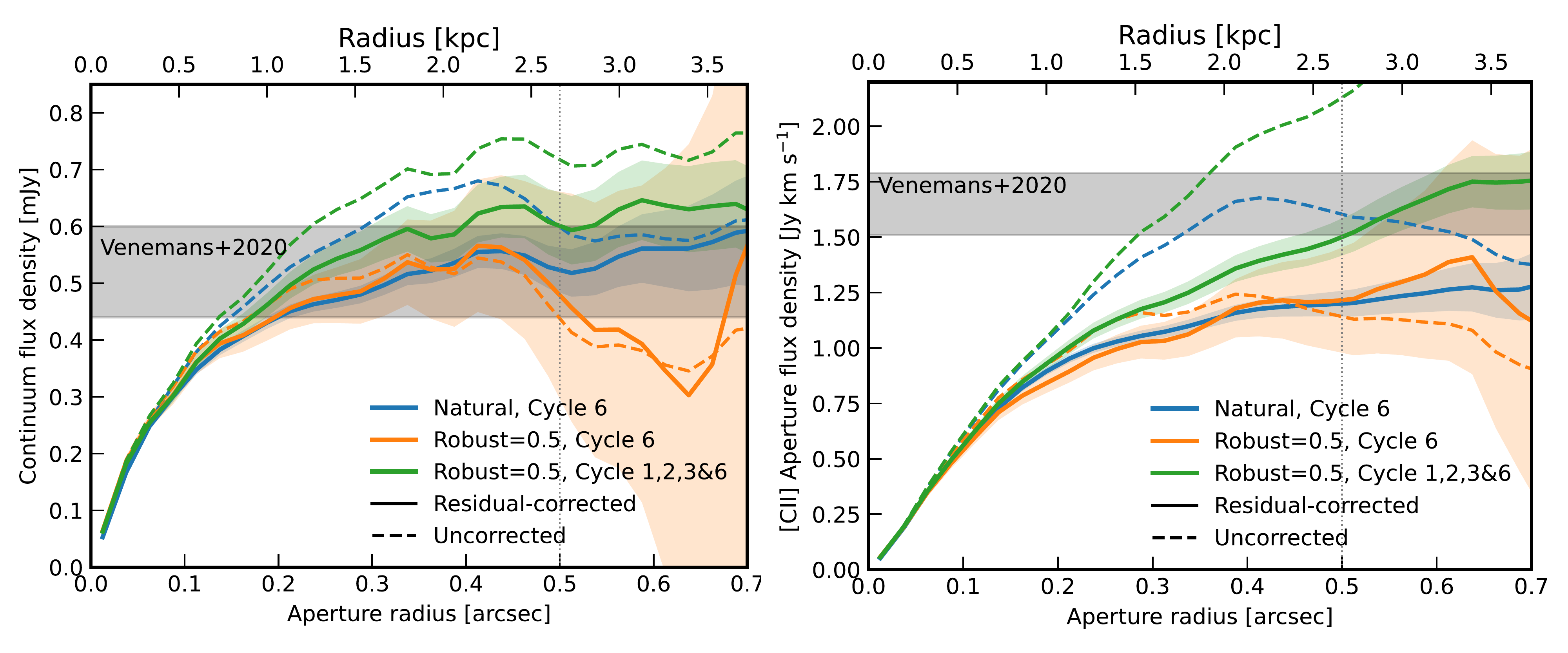}
    \caption{Curve of growth of the dust continuum (left) and [\ion{C}{2}] velocity--integrated emission (right). The flux density reported in \citet{Venemans2020} is recovered at an aperture size of $r=0\farcs5$ (e.g., $2 \rm{kpc}$ at $z=6.7$). The flux difference between the low- and high-resolution data is only $<10\%$ and the fluxes agree within the $1\sigma$ uncertainty, indicating that little or no diffuse emission ($\theta\gtrsim 0\farcs05$) is lost in the high-resolution data. We note that the errors shown here are only statistical (using the rms in the map and scaling for the aperture size) and do not include the $\sim10\%$ amplitude calibration uncertainty.
    }
    \label{fig:curve_of_growth}
\end{figure*}

The continuum was subtracted in the visibilities plane from the lower sideband containing the [\ion{C}{2}] emission using a first order polynomial fitted to the channels without emission ($|\nu -\nu_{\rm{CII}}|>2.5\times\ \rm{FWHM(\rm{CII})}$), where we used the conservative FWHM of $354\pm34$ from the low--resolution data \citep[][]{Neeleman2021}. The full data cube, the velocity--integrated [\ion{C}{2}] and the dust continuum images were imaged using \textit{multi--scale} cleaning down to $2\sigma$. The noise in the data cube is $\sigma = 61\ \mu \rm{Jy\ beam}^{-1}$ per $30$ MHz ($37\ \kms$) channel and $\sigma =  5.7\ \mu \rm{Jy\ beam}^{-1}$ in the dust continuum image. The velocity--integrated [\ion{C}{2}] image was created using the channels within $<1.25\times \rm{FWHM}$ of the center of the [\ion{C}{2}] line, and has an rms of $\sigma =  21\ \mu \rm{Jy\ beam}^{-1}$. Assuming a Gaussian line profile (which is verified in the data), such a map contains only $0.84$ of the flux and the [\ion{C}{2}] fluxes taken from the image are corrected accordingly. Additionally, all the fluxes and flux densities presented in this work include residual--scaling correction \citep[e.g.,][]{Jorsater1995,Walter1999,Novak2019}, as implemented in the latest version of \emph{interferopy} \citep*[v1.0.2,][]{interferopy}.

\section{The compact ISM of J0109--3047} \label{sec:ISM_results}

We show the dust continuum and velocity-integrated [\ion{C}{2}] imaging of J0109–3047 in Figure \ref{fig:maps}. We also show the curve of growth of the [\ion{C}{2}] and continuum fluxes in Figure \ref{fig:curve_of_growth}. Both the [\ion{C}{2}]  and dust emission are extremely compact, with most of the flux emitted within $r<0\farcs1$ ($\sim 500$ pc). The compactness of the source is not due to over-resolving as we recover within uncertainties the fluxes measured in lower-resolution observations \citep[][]{Venemans2016,Venemans2020}. Nonetheless, we still combine the archival lower-resolution observations with our $0\farcs04$ data to produce our final cubes and images in order to
capture any diffuse emission, and use $r = 0\farcs5$ apertures to capture the total fluxes.

We now proceed to measure the infrared luminosity assuming a common graybody dust emission assuming an opacity $\kappa_{\nu_{rest}} = \kappa_{\nu_0}(\nu_{rest}/\nu_0)^\beta$ with $\nu_0=c/(125\mu\rm{m})$ and $\kappa_0=2.64\ \rm{m}^{2}\ \rm{kg}^{-1}\ $ \citep{Dunne2003}. We assume $\beta=1.6$ and $T_{\rm{dust}}=47\ \rm{K}$ , values commonly consistent with the dust SED of most high-redshift quasars \citep[e.g.,][]{Bertoldi2003,Leipski2014,Venemans2016, Decarli2022}. We include CMB heating and contrast corrections as per \citet[][]{DaCunha2015}. Under these assumptions, the total continuum flux density extracted in an aperture of $r=0\farcs5$ is $f_\nu = 0.54 \pm 0.08\ \rm{mJy}$ can be extrapolated to a total infrared ($8-1000 \mu\rm{m}$) luminosity of $L_{\rm{TIR}}=1.8\ \times10^{12}\ L_\odot$\footnote{Note that the error on the dust mass only reflects the formal error on the continuum flux density. The dust temperature and optical depth can deviate from the assumption of $T=47\, \rm{K}$ \citep[e.g.,][]{Walter2022} and optical thin emission \citep[e.g.,][ for high--redshift SMG studies]{Riechers2013, Spilker2014} which can lead to a systematic uncertainty of factor 2-3 on the total infrared luminosity, dust mass and SFR \citep[see, e.g.,  ][]{Venemans2018}}. Converting this to a star--formation rate (SFR) estimate using the conversion in \citet{Kennicutt2012} yields a total SFR of $\simeq 180\ M_\odot \rm{yr}^{-1}$. The total dust mass is $(5.2\pm0.9) \times 10^{7}\ M_\odot$ , which, assuming a fixed dust--to--gas ratio of $1/100$, yields a total gas mass of $(5.2\pm0.9) \times 10^{9} M_\odot$. 

In the absence of multi-frequency observations sampling the FIR dust SED of J0109--3047, the total dust mass is degenerate with the dust temperature and can vary by an order of magnitude. Additionally, the gas-to-dust ratio (GDR) in this object is unknown, although previous studies have shown that similar high-redshift quasars have GDR=70-100 \citep[e.g.][]{Riechers2013,Wang2016, Decarli2022}. Previous CO observations of our target independently constrain the total molecular gas mass of J0109--3047 to $(1.0\pm0.2) \times 10^{10} M_\odot$ \citep[][]{Venemans2017b}, assuming a CO conversion factor $\alpha_{CO}=0.8\ M_\odot (\rm{K\ km\ s^{-1}})^{-1}$ and CO SLED typical of ultra-luminous infrared galaxies and high-redshift quasars \citep[e.g.][]{Bolatto2013,Carilli2013}. Assuming a GDR of 50(150), and a molecular gas fraction of $0.75$ this translates to a total dust mass of $2.67\pm0.54 \ (0.89\pm0.18) \times 10^{8}\ M_\odot$. Thus the dust and gas mass estimates used in this paper ($M_d = (5.2\pm0.9) \times 10^{7}\ M_\odot$, $M_{\rm{gas}} = (5.2\pm0.9) \times 10^{9} M_\odot$ ), even accounting for large uncertainties in the various conversion factors, are therefore at the least massive end of the range allowed by the CO measurements.

The central beam (corresponding to a region with an effective radius $r=144\ \rm{pc}$) accounts for $25\pm3 \%$ of the total dust emission and thus the total FIR luminosity. Whilst the relative concentration of continuum flux density in the central hundred parsecs is similar to that observed in J2348--3054 \citep[][]{Walter2022}, the infrared surface brightness and SFR surface densities is lower by a factor $\sim 30$ ($\Sigma_{\rm{TIR}}=7.3\times10^{12}\  L_\odot\ \rm{kpc}^{-2},  \Sigma_{\rm{SFR}} \simeq 730\ M_\odot \rm{yr}^{-1} \rm{kpc}^{-2}$ in the central beam). This difference is driven, in part, by the lower continuum emission, but mainly by the different dust temperatures. Unlike J2348--3054, the continuum flux densities in the central region of J0109--3047 are low enough to proceed under that assumption of optically thin emission and using a temperature $T_d=47\, \rm{K}$. 

\begin{figure}
    \centering
    \includegraphics[width=0.5\textwidth]{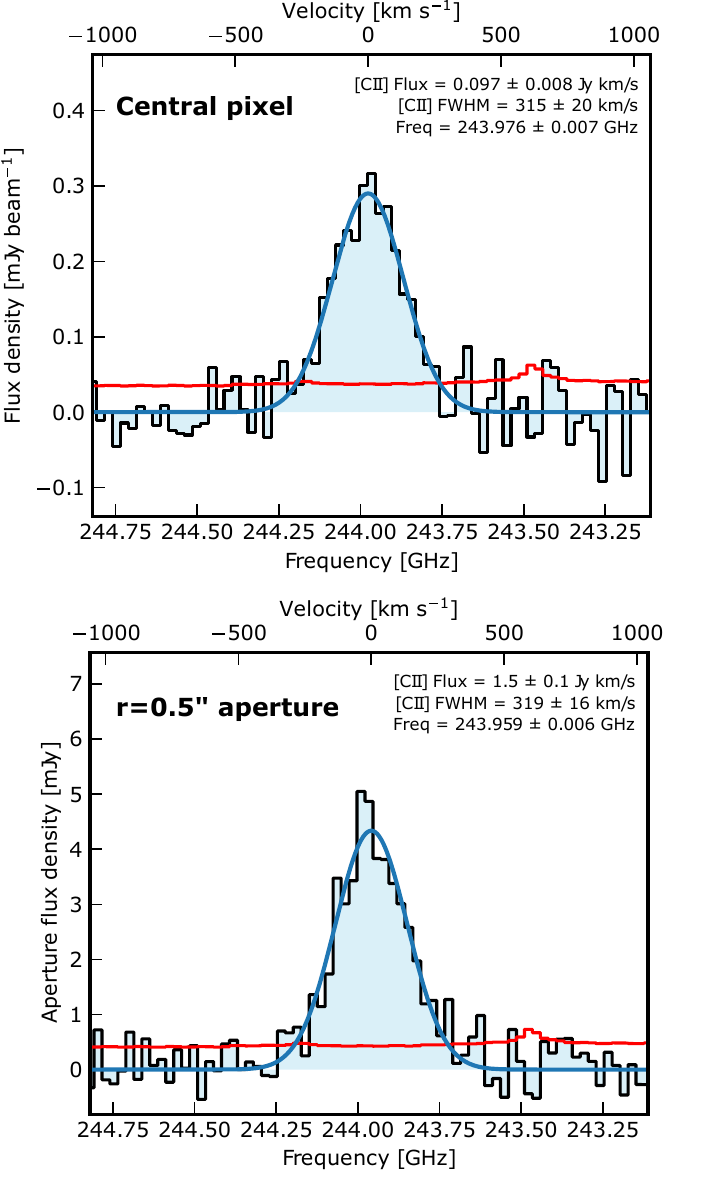}
    \caption{ [\ion{C}{2}] spectrum of J0109--3047 in the central pixel (defined as the FIR peak, upper panel) and the optimal $r=0.5"$ aperture (lower panel), using the combined new high-resolution data and archival low-resolution observations. The error per frequency channel (defined as the $\sigma$-clipped rms per pixel in each channel, and scaled by the square root of the number of pixels in the aperture) is shown in red, a single Gaussian fit is overplotted in blue. The best-fit parameters and errors of the Gaussian fit are indicated in the top right corners. The frequency, amplitude and FWHM of the [\ion{C}{2}] line is consistent with that of the previous observations, and we find no evidence for an additional broad component in the [\ion{C}{2}] emission line profile.}
    \label{fig:cii_spec}
\end{figure}

The aperture--integrated ($r=0\farcs5$) [\ion{C}{2}] line central frequency is $\nu_{\rm{CII}}=243.959\pm 0.006\ \rm{GHz}$ (implying a redshift $z_{\rm{CII}}=6.79039 \pm 0.00019 $), the FWHM is $319\pm 16 \, \kms{}$ and the line luminosity is $(2.44\pm0.23) \times 10^{9}\, L_\odot$, in good agreement with the values determined from the low-resolution data \citep[][]{Venemans2020}. As in \citet[][]{Walter2022}, we find that the [\ion{C}{2}] emission is less concentrated than the continuum ($8\%$ vs $25$\% in the central resolution element). We do not find any evidence for a deviation from a Gaussian line profile in the [\ion{C}{2}] line (see Fig. \ref{fig:cii_spec}). Such a deviation could indicate an ongoing or recent merger \citep[e.g.,][]{Venemans2019, Decarli2019} or an outflow \citep[e.g.,][]{Maiolino2005, Novak2020}. We also do not find any [\ion{C}{2}] companions in the high-resolution data. We summarise the properties of the observed continuum and [\ion{C}{2}] emission as well as their derived properties in Table \ref{tab:properties}.

\begin{table}
    \centering
    \begin{tabular}{c|c|c}
         & Central beam &
         $0\farcs5$ aperture   \\
        & ($r=144$ pc)&($r=1.9$ kpc) \\ \hline
        
        $f_{c}\ \rm{[mJy]} $  & $0.134 \pm 0.006$ & $0.536 \pm 0.075$ \\
        $F_{\rm{CII}}\ \jykmsmath $  & $0.114 \pm 0.009$ & $1.411 \pm 0.079$ \\
        $L_{\rm{TIR}}^{a} [10^{12} L_\odot]$ & $0.45\pm0.02$ & $1.8\pm0.3$ \\
        $L_{\rm{CII}}  [10^{9} L_\odot]$ & $0.26 \pm 0.02$ & $2.44\pm 0.23$ \\
        $M_d^{a} [10^{7} M_\odot] $ & $1.3\pm0.1$ & $5.2\pm0.9$ \\ 
        SFR(TIR)$^{b} [M_\odot \rm{yr}^{-1}]$ & $45\pm2$ & $180\pm30$ \\
        SFR([\ion{C}{2}])$^{c} [M_\odot \rm{yr}^{-1}]$ & $77\pm6(^{+39}_{-38})$ & $531\pm21(^{+528}_{-265})$ \\
    \end{tabular}
    \caption{Properties of J0109--3047 derived from our high-resolution data.$^{a}$ TIR luminosities and dust masses are derived assuming a dust temperature $T_d=47\ \rm{K}$ and spectral index $\beta=1.6$. Consequently, the error only include the uncertainty on the single dust continuum measurement and not the assumed dust parameters. $^{b}$ SFR are computed from the TIR luminosity using the \citet[][]{Kennicutt2012} relationship.  $^{c}$ SFR calculated using the \citet[][]{Herrera-Camus2018} “High-$\Sigma_{\rm{TIR}}$" scaling relation. The second uncertainty in parenthesis derives from the scaling relation calibration uncertainty.     \label{tab:properties}} 
\end{table}

The most striking feature of Fig.~\ref{fig:maps} might be the apparent offset between the near--infrared (NIR) position of the quasar \citep[as derived from the GAIA--corrected position of stars in the near--infrared imaging of the quasar fields][]{Venemans2020} and that of the FIR continuum and [\ion{C}{2}] emission line. We will revisit the possibility that the offset between the quasar and its host galaxy is real in Section \ref{sec:offset}, after having detailed our constraints on the black hole mass from the [\ion{C}{2}] kinematics.

\section{The dispersion--dominated [\ion{C}{2}] kinematics of J0109--3047} \label{sec:kinematics}

We show the [\ion{C}{2}] kinematics in the bottom row of Figure \ref{fig:maps}. The mean velocity (“moment 1") and velocity dispersion (“moment 2") fields are constructed by fitting a Gaussian profile to each pixel in the datacube using \emph{QUBEFit} \citep{qubefit}. The [\ion{C}{2}] kinematics of J0109--3047 are characterized by the absence of ordered rotation and a constant high velocity dispersion. We fit the [\ion{C}{2}] intensity profile and kinematics with a spherically symmetric dispersion--dominated model with an exponentially declining intensity profile and a constant velocity dispersion \citep[e.g.,][ and see Appendix \ref{app:models} for the best-fit model and residuals]{Neeleman2021}. The best-fit model has a constant velocity dispersion of $\sigma_v = (137 \pm 6)\ \kms{}$. We find no evidence for an exponentially declining velocity profile (the Bayesian Information Criterion \citep[BIC, e.g.,][]{Kass1995} is increased by $\Delta \rm{BIC} = 12.1$).

The lack of a clear velocity gradient could alternatively be explained by a rotating disk being observed face--on. We disfavor this scenario \textit{a priori} based on previous observations of $z>6$ quasar hosts. \citet{Pensabene2020,Neeleman2021} showed that $z>6$ quasar host galaxies kinematics are divided equally between mergers, ordered rotation and an absence of ordered rotation. The wide range of inclinations found in the objects with ordered rotation ($20\lesssim i \lesssim 90$) indicates that the host galaxies are not always seen face--on. If the third of $z>6$ quasar hosts without a clear velocity gradient were seen face--on, one would expect a smaller velocity dispersion as the beam probes a smaller region of the galaxy due to the inclination. On the contrary however, quasar host galaxies without ordered rotation (including J0109--3047) have significantly higher velocity dispersion than those where ordered rotation is detected \citep[][]{Neeleman2021}, suggesting they have more turbulent kinematics and are ´dispersion--dominated' \footnote{It is worth noting that in the AGN unification model Type 1 quasars (e.g. such as those found at $z>6$) are supposed to be seen face-on. Therefore, the recent results on the quasar host kinematics discussed here imply that the SMBH accretion disk is not always aligned with the host galaxy rotation in $z>6$ luminous quasars.}.

We fit the kinematics of J0109--3047 with a thin disk model in \textit{Qubefit} to understand whether the absence of rotation could be produced by a face-on disk. We find that the inclination is constrained to $i=(50.9^{+1.7}_{-2.2}) ^{\circ}$, in agreement with the similar approach of \citet{Neeleman2021} using $0.2"$-resolution data. The inclination is mainly constrained by the elongation of the [\ion{C}{2}] emission. Indeed, a fit to [\ion{C}{2}] emission using CASA's \textit{imfit} task gives a best-fit deconvolved major axis $a=200\pm24\ \rm{mas}$, minor axis $b=126\pm15\ \rm{mas}$, and position angle $\rm{PA}=55\pm10\ \rm{deg}$,  implying an inclination of $(50.9\pm0.1)^{\circ}$. J0109-3047 is thus not seen completely face-on and the inferred maximum rotational velocity is $<4.3\,  \kms{}$ at the $2\sigma$ level, which is consistent with the absence of any velocity gradient in the data (e.g. Fig. \ref{fig:maps}). The kinematics in J0109--3047 are thus ´dispersion--dominated' with a $v_{\rm{rot}} / \sigma_v <0.02 (2\sigma)$ largely below the formal cutoff of $1-3$, \citep[][]{Epinat2009, Burkert2010,ForsterSchreiber2018}. Interpreting the kinematics with a thin disk model, or a model showing ordered rotation is thus improper and we base our conclusions on ´dispersion-dominated' models in this work.

Finally, we follow \citet[][]{Neeleman2021} in deriving a dynamical mass using the best-fit [CII] scale length, the rotational velocity and the velocity dispersion. We find a total dynamical mass $M_{\rm{dyn}} =[2.34_{-0.08}^{+0.14} (\pm0.94)] \, \times 10^{10}\, M_\odot $, in good agreement with the \citet[][]{Neeleman2021}{}{} value $M_{\rm{dyn}} = [1.89^{+0.24}_{-0.27}(^{+1.1}_{-1.3}) ] \times 10^{10} \ M_\odot$.

\begin{figure*}
    \includegraphics[width=\textwidth, trim = 1.0cm 0 1.5cm 0,clip]{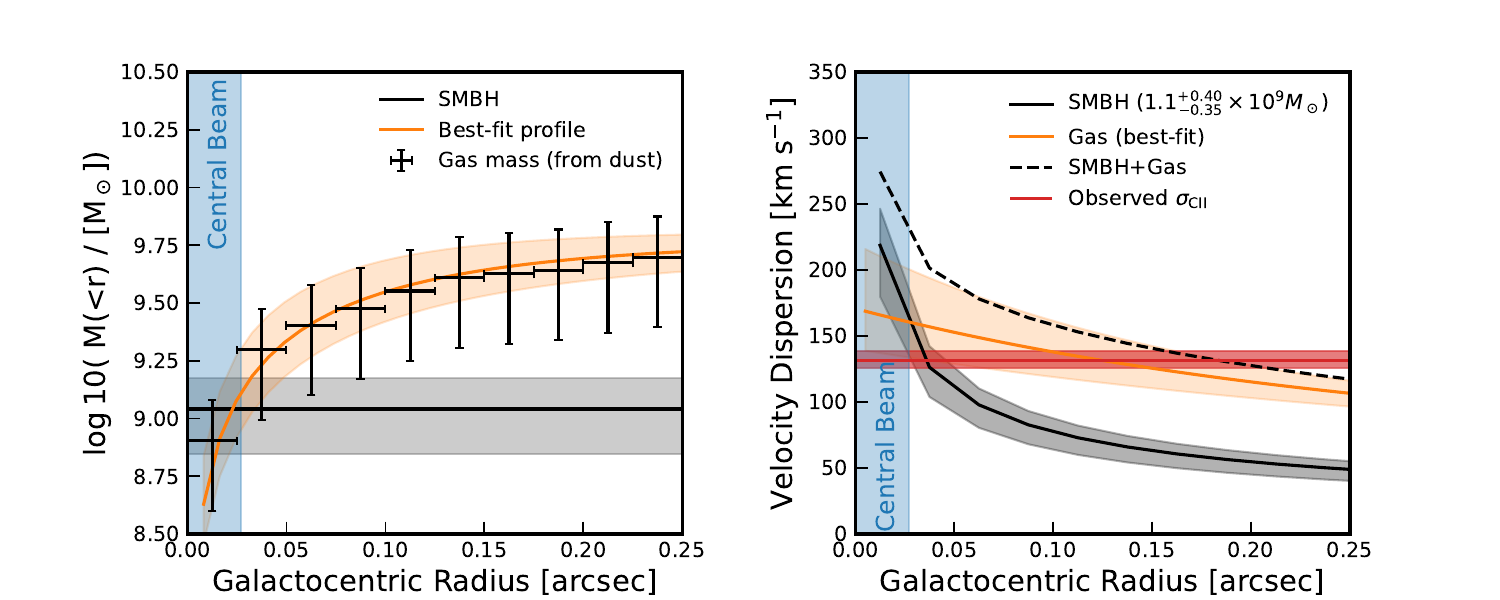}
    \caption{Integrated mass (left) and velocity (dispersion) profiles as a function of galactocentric radius for the SMBH and gas component of the galaxy. The integrated gas mass profile is derived from the radially--averaged dust profile assuming a gas--to--dust ratio of $100\pm50$. The velocity dispersion is calculated by approximating the gas as virialized. To first order, the velocity dispersion derived from the gas mass is consistent with that observed and is constant.}
    \label{fig:masses_velocities}
\end{figure*}

The absence of ordered rotation in a quasar host galaxy at $z>6$ is somewhat surprising. The mass of this $z=6.79$ black hole is $M_{\rm{BH}} = (0.6-1.1) \times 10^{9} M_\odot$ \citep[][]{Farina2022, Yang2023}. This implies that it has been accreting at a fraction of the Eddington rate for a sizeable time, and that the accretion disk and likely the host galaxy disk necessary to funnel the gas have been present for a similarly long time. Even in the scenario with the shortest growth time (e.g., assuming a $10^5$ M$_\odot$ seed direct collapse black hole, $\lambda_{\rm{Edd}}=1$, a duty cycle of 1 and a Salpeter time of $45\ \rm{Myr}$), the BH must have been accreting for $\sim 300\ \rm{Myr}$ \citep[e.g.,][]{Banados2018,Wang2021}, whereas in the slowest growth scenarios the SMBH must have been accreting at $0.2-0.5$ Eddington rate for at least $600-700$ Myr. The absence of ordered rotation, which is expected in simulations up to $z\sim 8$ \citep[e.g.,][]{Pillepich2019}, could suggest that the system is in a post--merger phase with dispersion--dominated kinematics. However, we do not find evidence for such a recent merger, whether as a faint companion or as a gradient in the velocity field. We can speculate that the merging galaxy would have to be a low-mass galaxy that is sufficiently faint to leave no trace in the observed [\ion{C}{2}] data (Fig.~\ref{fig:maps}), but it would need to be sufficiently large to completely disturb any existing rotation.

\section{The sphere of influence of the SMBH} \label{sec:soi}

Assuming that the SMBH of J0109--3047 is located at the peak of its dust continuum (and [\ion{C}{2}] line) emission, we can place interesting constraints on the mass of the SMBH. As discussed in Section \ref{sec:ISM_results}, under standard assumptions about the dust properties the dust mass in the central pixel is $M_d =(1.3\pm0.2) \times 10^{7} M_\odot$. Assuming a gas--to--dust ratio of $100$, this implies that the gas mass inside the central beam is $M_{gas} \simeq 1.3 \times 10^{9} M_\odot$, which is comparable to the black hole mass estimates from the rest--frame UV $M_{\rm{BH}} = (1.1\pm0.4) \times 10^{9} M_\odot$ \citep[][]{Farina2022} and rest-frame optical $M_{\rm{BH}} = (0.60\pm0.05) \times 10^{9}\ M_\odot$ \citep[][]{Yang2023}. Therefore the sphere of influence (e.g., the region where the BH dominates the gravitational potential) corresponds, to first order, to the size of the central beam ($r=144\ \rm{pc}$), where we should be able to detect the imprint the of SMBH on the gas velocity dispersion \citep[e.g.,][]{Lupi2019}.

Since the galaxy is dispersion--dominated, the impact of the SMBH on the kinematics would only be visible in the velocity dispersion field. Using the virial theorem, the [\ion{C}{2}] dispersion in a pixel at distance $r$ from a SMBH of mass $M_{\rm{BH}}$ is \citep[e.g.,][]{Decarli2018, Lupi2019}
\begin{equation}
    \sigma_{\rm{CII}, SMBH} = \sqrt{\frac{2}{3}\frac{GM_{\rm{BH}}}{r} }  \text{\,\,\,\, .}
    \label{eq:sigma}
\end{equation}
For the central brightest resolution element, the velocity dispersion is $\sigma_{\rm{CII}} =  (134\pm8)\ \kms$, which yields a dynamical mass of $(0.87\pm0.11) \times 10^{9} M_\odot$. Although this is compatible with the SMBH mass measured from rest--frame UV/optical ( $(1.1\pm0.4) \times 10^{9} M_\odot$ /  $(0.60\pm0.05) \times 10^{9}\ M_\odot$ ) and the gas mass estimate ($M_{gas} \simeq 1.3 \times 10^{9} M_\odot$), the dynamical mass is only about $50\%$ of the combined SMBH and gas mass in the central $144\ \rm{pc}$. 

We show the velocity dispersion radial profile expected assuming the gravitational potential of a $(1.1\pm0.4) \times 10^{9} M_\odot$ SMBH, the gas mass derived from the dust profile, or both, on Figure \ref{fig:masses_velocities}. The observed velocity dispersion $\sigma_{\rm{CII}} =  (134\pm8)\ \kms$ is incompatible with the sole BH or BH+gas radial velocity dispersion profiles, but consistent with a model where only the gravitational potential of the gas is considered. 

 \begin{figure*}
     \centering
     \includegraphics[width=0.9\textwidth]{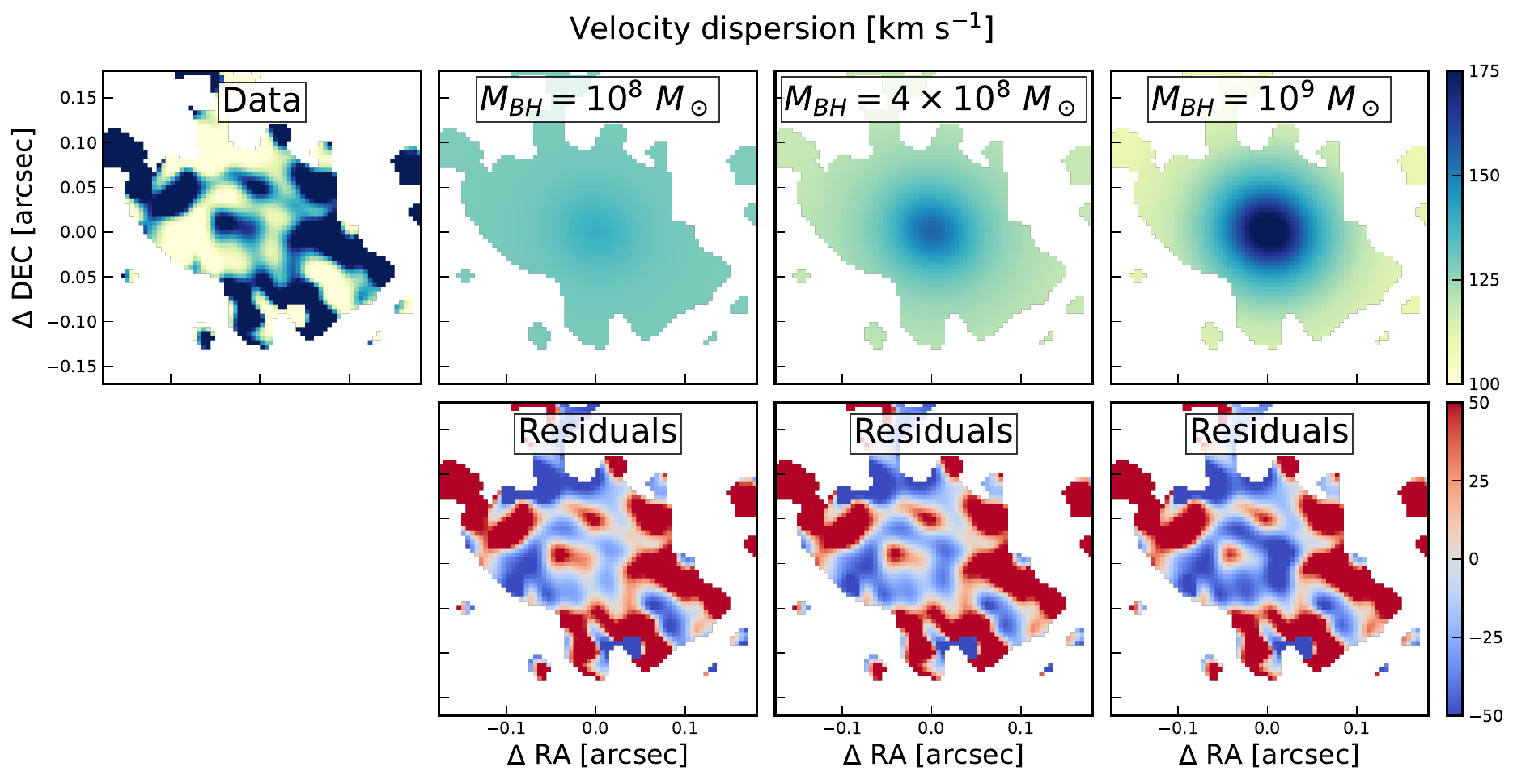}
     \caption{ Velocity dispersion of the [\ion{C}{2}] line in J0109--3047 (top left) and best-fit models for given black hole masses between $10^{8}-10^{9}\ M_\odot$ (first row). The residuals of the models are shown in the second row. The data, models and residuals are only shown for pixel in which the velocity--integrated [\ion{C}{2}] emission exceeds $3\sigma$ (as in Figure \ref{fig:maps}). The reduced $\chi^2$ statistic (adjusted for the kernel size) for the three models are, in order, $\chi^2_{red}=1.097,1.097,1.101$. }
     \label{fig:BH_dispersion_models}
 \end{figure*}

To investigate the impact of the SMBH on the kinematics of J0109--3047 further, we construct a simple “dispersion--dominated + SMBH" model in \emph{QUBEFit}. In this model the velocity dispersion is the sum of the SMBH component of Eq.~\ref{eq:sigma} and a constant dispersion value throughout the quasar host ($\sigma_{\rm{CII, tot}}^2 =  \sigma_{\rm{CII}, SMBH}(r)^2 + \sigma_{\rm{const}}^2$). The intensity profile is assumed to be exponentially declining as in the constant dispersion model (see Section \ref{sec:ISM_results}). We note that any additional contribution (besides the central SMBH) to the kinematics as a constant is in agreement with the inferred gas mass profile (see Fig. \ref{fig:masses_velocities}).

In Fig. \ref{fig:BH_dispersion_models} we show the best-fit dispersion fields for different SMBH masses from $10^{8}\ M_\odot$ to $10^{9}\ M_\odot$. We find that the [\ion{C}{2}] kinematics of J0109--3047 are clearly incompatible with a $\sim 10^{9}\ M_\odot$ SMBH. A full MCMC fit of the model to the data yields an upper limit of $M_{\rm{SMBH}} < 6.5 \times 10^{8}\ M_\odot (2\sigma)$ (see Appendix \ref{app:models} for the full posterior distribution of the model parameters). However, even for the maximum-likelihood model ($M_{\rm{BH}} = 2.4\times 10^{8}\ M_\odot$), the BIC is slightly higher than for a model without a black hole ($\Delta \rm{BIC}= 5.2$), showing that any SMBH contribution to the dispersion velocity field is disfavored by the observations. 

One way to alleviate the tension with the rest-frame UV mass measurement could be to change the radial [\ion{C}{2}]--emitting gas profile. By definition, the observed [\ion{C}{2}] kinematics are a luminosity--weighted, beam--convolved realisation of the intrinsic kinematics. Following Eq. \ref{eq:sigma}, the velocity dispersion increases exponentially close to the black hole, and due to the exponential intensity profile, these inner regions will contribute more to the beam--convolved velocity dispersion measurement in the center. As a result, the observed velocity dispersion could be reduced, if the [\ion{C}{2}] intensity profile is not increasing close to the black hole (for example due to feedback).

To address this further, we have used a toy model where the gas density profile follows an exponentially declining profile with a central gap where the [\ion{C}{2}] emission is null. As in the fiducial model, the velocity dispersion is composed of the SMBH component and a constant. We use this simple model to calculate the size of the central gap necessary to “hide" the SMBH impact on the [\ion{C}{2}] kinematics tracer. We find that, for a SMBH with a fixed mass $M_{\rm{BH}}=1.1\times10^9 M_\odot$, the best-fit central gap is constrained to be $r< 22\ \rm{pc}$ ($2 \sigma$) to reproduce the [\ion{C}{2}] profile and kinematics. The best-fit model has $r_{\rm{pc}}=0.015^{+0.015}_{-0.010}\ \rm{pc}$ (see Appendix \ref{app:models}), and is formally ruled out with an increased $\Delta \rm{BIC} = 10.42$ compared to the model without a gap. Moreover, a central gap in the gas distribution would be at odds with simulations and observations where the central $\sim 400-500\ \rm{pc}$ region contains up to $\sim 10$ times the mass of the BH in gas \citep[e.g.,][]{Lupi2022,Walter2022}. 

In summary, the flat velocity dispersion profile implies a flat radial mass density profile. The constant dispersion implies that the underlying mass distribution is not centrally peaked, consistent with the expectations of the gas mass distribution derived from the far-infrared continuum emission under standard assumptions. This leaves only few alternatives to explain the absence of a central peak in the velocity dispersion. One possibility is that the gas mass decreases in the central $200$ pc in order to compensate the presence of a $0.6-1.1\times10^{9}\ M_\odot$ black hole and produce a flat mass profile. However, we have previously excluded the presence of a central gap in the [\ion{C}{2}]--emitting gas, and the FIR continuum shows no sign of a central gap either. A decrease in the central gas mass would imply fine-tuning of the physical properties of the ISM at the center of J0109–3047 (to `offset' the mass of the SMBH). If such a conspiracy is excluded, the black hole mass is either smaller than expected, as discussed in this section, or the black hole is not located at the center of the galaxy as traced by the dust continuum, as discussed in Section \ref{sec:offset}.

If confirmed, and applicable to the larger population of $z>6$ luminous quasars, a mass of $\sim 10^{8} M_\odot$ for the SMBH in J0109--3047 would have several interesting implications for early SMBH growth and formation. First, it would alleviate the need for massive seeds and/or super--Eddington accretion events at $z>7$ \citep[e.g.,][]{Banados2018,Wang2021,Volonteri2021}. Second, a SMBH mass of $10^8 M_\odot$ for a total galaxy (dynamical) mass of $2.34 \times 10^{10} \, M_\odot$ (see Section \ref{sec:kinematics}) would place it on the local relation, meaning that J0109--3047 is not part of an overmassive SMBH population at $z>6$ \citep[][]{Pensabene2020,Neeleman2021}, and the offset from the local relation seen in the $z>6$ luminous quasar sample could be due to systematic overestimation of black hole masses. Third, the accreting BH at the heart of J0109--3047 would be definitive evidence for super--Eddington accretion at $z>6$ with an Eddington ratio $\lambda_{\rm{Edd}} \gtrsim 5$. 

\section{An offset or recoiling SMBH at redshift z=6.79?}
\label{sec:offset}
The discussions in Section \ref{sec:soi} relied on the assumption that the black hole is located at the center of the host FIR continuum emission. However, if the accreting SMBH is not located at the center of the host galaxy, the [\ion{C}{2}] kinematics are not expected to be strongly influenced by the SMBH, and the tension between the dynamical mass derived in Section \ref{sec:kinematics} and the one derived from the rest-frame UV \ion{Mg}{2} line \citep[][]{Farina2022} would disappear.

\citet[][]{Venemans2020} already reported a $\sim 2\sigma$ offset between the GAIA--corrected NIR position of the quasar and that of the ALMA continuum of the host imaged with a $0\farcs20 \times 0\farcs17$ beam. At the time, the error budget on the NIR--mm offset was dominated by the ALMA continuum (the quasar NIR position error is $0\farcs02\times0\farcs05$), and \citet{Venemans2020} cautiously considered the offset as a $2\sigma$ outlier in their larger sample of $27$ quasars, and warned that the uncertainties on the NIR position could be underestimated.

However, our new high-resolution data drastically reduces the error on the ALMA continuum position. Using the CASA task \textit{imfit}, we fit a 2D elliptical Gaussian to the $\sim$250 GHz continuum. The best--fit ellipse is centered at 01:09:53.12534 -30:47:26.32077 with an error of $0\farcs00016\times 0\farcs00162$, respectively. The major/minor axis and flux of the Gaussian are well constrained and the residuals of the model are negligible. Fitting in the UV plane (CASA task \textit{uvmodelfit}) gives an identical value with a best-fit position 01:09:53.12534 -30:47:26.32077 and error $0\farcs00016\times 0\farcs00162$. For our high-resolution observations, the astrometric uncertainty on the calibrator ranges between 1-7 mas between the tracks, for an average rms of $\sim 3\ \rm{mas}$, consistent with the uncertainty on the 2D Gaussian fit. Following the same steps, the ALMA ‘check' source (J0115--2804) position is consistent with its radio position within the uncertainty, confirming the absence of any phase calibration issues that could affect the position of J0109--3047.

Our new 300--pc resolution thus confirms the offset between the NIR position of the quasar and the FIR host galaxy continuum (see Fig. \ref{fig:maps}). Similar offsets between two faint $z\sim 6$ quasars and their host galaxies were recently reported in \textit{JWST} NIRCam observations \citep[][]{Ding2022b}. This suggest that the offset in J0109--3047 might be real, and not a statistical outlier in a larger sample. Additionally, J0109--3047 shows extremely blueshifted \ion{C}{4} $1549\ \rm{\AA}$ ($-4573^{+304}_{-293}\ \kms{}$) and \ion{Mg}{2} $2800\ \rm{\AA}$ ($-1009^{+371}_{-426}\ \kms{}$) with respect to the host galaxy [\ion{C}{2}] 158 $\mu \rm{m}$ line compared to other $z\sim6$ luminous quasars \citep[][]{Schindler2020}. The [\ion{C}{2}] redshift is coincident with the start of the neutral IGM Lyman-$\alpha$ absorption in the quasar spectrum \citep[][]{Venemans2013}, confirming that the \ion{Mg}{2} and \ion{C}{4} emission lines are indeed offset with respect to the host galaxy. Both the spatial and the velocity offset are compatible with that expected for rapidly recoiling black holes following a major merger with misaligned spins \citep[e.g.,][]{Campanelli2007,Blecha2016}. The high velocity dispersion in the ISM would be the signature of a past merger which would have triggered the ejection of the SMBH out to $\sim 0.8\ \rm{kpc}$ from the center as traced by the FIR emission. 

We caution that this hypothesis must be confirmed with a more precise measurement of the quasar position in the NIR. Indeed, the PSF width and pixel size of the VIKING imaging used by \citet[][]{Venemans2020} is suboptimal to determine a position at the $\sim 10\ \rm{mas}$ level required. JWST/HST is unlikely to provide much improvement as the typical number of (unsaturated) GAIA stars in their field of view is extremely low. This in turn limits the ability of such observations to correct the astrometry to the GAIA/ALMA frame. What is required is wide field--of--view, high--sensitivity ground--based imaging with AO such as HAWK-I on the VLT that can provide both the small PSF, high spatial pixel sampling and large number of GAIA stars per image necessary for this measurement.

\section{Conclusion} \label{sec:conclusions}
We have presented high--resolution ($300$ pc) [\ion{C}{2}] observations of the $z=6.79$ quasar J0109--3047 with ALMA. This data, combined with previous short baseline observations in Cycle 1, 2 and 3, provide intriguing constraints on the ISM of J0109--3047 and its central SMBH. In particular, we report the following findings:
\begin{itemize}
    \item The FIR continuum and [\ion{C}{2}] emission is extremely compact, with most of the flux arising within $\lesssim 500\ \rm{pc}$ of the core of the galaxy, and a maximal extension up to $\sim 2\ \rm{kpc}$. $25\%$ ($8\%$) of the continuum ([\ion{C}{2}]) emission emanates from the central beam, which corresponds to a region with an effective radius $r=144\ \rm{pc}$. The small size and compactness is similar to what is found for the $z=6.9$ quasar J2348--3054 \citep{Walter2022} and lower-redshift ALPINE galaxies \citep[][]{Jones2021a}. 
    \item The [\ion{C}{2}]--based kinematics do not show evidence for ordered rotation, and are instead dispersion--dominated, as was already suggested by the available $0.2"$-resolution data \citep[][]{Neeleman2021}. The best-fit velocity dispersion is high ($\sigma_{\rm{CII}}=  (137\pm6) \ \kms$) and does not change as a function of galactic radius. We show the [\ion{C}{2}] kinematics are inconsistent with a face-on disk. The absence of ordered rotation could stem from a recent merger event, although we do not find any spectral or spatial signature of a merger or a companion in the [\ion{C}{2}] emission.
    \item Taking into account the gas mass in the center of the galaxy, we find that for a black hole mass of $M_{\rm{BH}}=1.1\times10^{9}$ / $M_{\rm{BH}}=0.60 \times10^{9}$, as derived from the rest-frame UV/optical lines of the broad line region \citep[][]{Farina2022, Yang2023}, the black hole sphere of influence is comparable to the resolution of the observations. Under standard assumptions, we show that the dynamical mass in the central pixel is twice lower than the combined gas and BH masses, implying that either [\ion{C}{2}] is poor tracer of the kinematics close to the BH, the gas mass is overestimated, or the BH mass is underestimated. The constant flat velocity dispersion implies a flat mass profile consistent with the gas mass derived from the dust continuum emission map. Under the assumption that the properties of the ISM do not significantly change at the center of J0109--3047, our kinematical analysis puts a conservative $2\sigma$ upper limit on a central SMBH mass of $M_{\rm{BH}} < 6.5\times 10^{8} M_\odot$.
    \item Alternatively, the accreting SMBH/quasar might not be located at the center of its host galaxy, as suggested by the apparent offset between the NIR position of the quasar and our ALMA high-resolution imaging of the host galaxy. This would explain the absence of any velocity dispersion increase in the [\ion{C}{2}] kinematics and suggest that the SMBH in J0109--3047 is offset, or recoiling, probably due to a past merger event.
\end{itemize}

Future JWST spectroscopy of the rest-frame optical emission lines might determine whether the masses determined from the \ion{Mg}{2} lines of the broad line region are to be trusted at these high redshifts and luminosities \citep[see e.g.][, for an early comparison of \ion{Mg}{2} and $H\beta$-based masses]{Yang2023}. Meanwhile, further ALMA hyper--resolution observations of additional $z>6$ quasar host galaxies are crucial to reveal galaxy-black holes spatial offsets and/or undermassive black holes are prevalent in the $z>6$ population, and study their resolved ISM. In either case, the consequences on early SMBH growth and SMBH--galaxy co--evolution models are far reaching.

\begin{acknowledgments}
The authors thank the anonymous referee for comments and suggestions which improved this paper. RAM, MN, FW acknowledge support from the ERC Advanced Grant 740246 (Cosmic\_Gas). RAM acknowledges support from the Swiss National Science Foundation (SNSF) through project grant 200020\_207349.

This paper makes use of the following ALMA data: ADS/JAO.ALMA\#2013.1.00273.S,\#2015.1.00399.S, \#2021.1.00800.S. ALMA is a partnership of ESO (representing its member states), NSF (USA) and NINS (Japan), together with NRC (Canada), MOST and ASIAA (Taiwan), and KASI (Republic of Korea), in cooperation with the Republic of Chile. The Joint ALMA Observatory is operated by ESO, AUI/NRAO and NAOJ. 
\end{acknowledgments}

\vspace{5mm}
\facilities{ALMA}

\software{astropy \citep{TheAstropyCollaboration2018}, 
numpy \citep{Numpy2020}, matplotlib \citep{Hunter2007}, scipy \citep{Virtanen2020}, QubeFit \citep{qubefit}, Interferopy \citep*{interferopy}}, emcee \citep[][]{Foreman-Mackey2013} 

\bibliography{library}{}
\bibliographystyle{aasjournal}

\appendix

\section{Kinematic models and full parameter posterior distributions} \label{app:models}
\begin{figure*}[h!]
    \centering
    \includegraphics[width=0.8\textwidth]{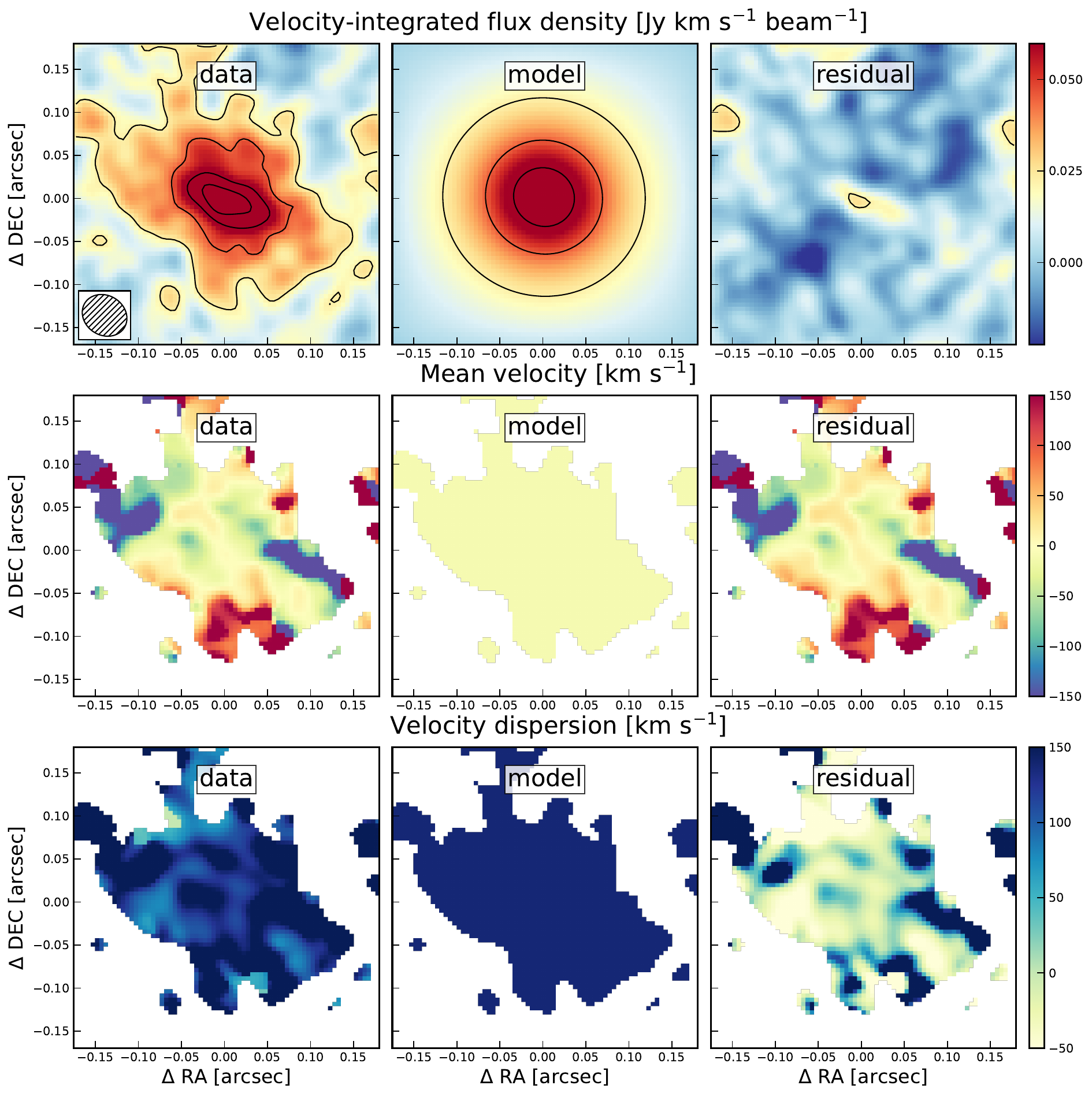}
    \caption{Comparison between the observed [\ion{C}{2}] emission and the best-fit dispersion--dominated model. This spherically symmetric model has an exponentially declining intensity profile and a constant dispersion velocity. The contours are start $3\sigma$ and increase in steps of 3.}
    \label{fig:Bulge_noBH}
\end{figure*}

\begin{figure*}
    \centering
    \includegraphics[width=0.6\textwidth]{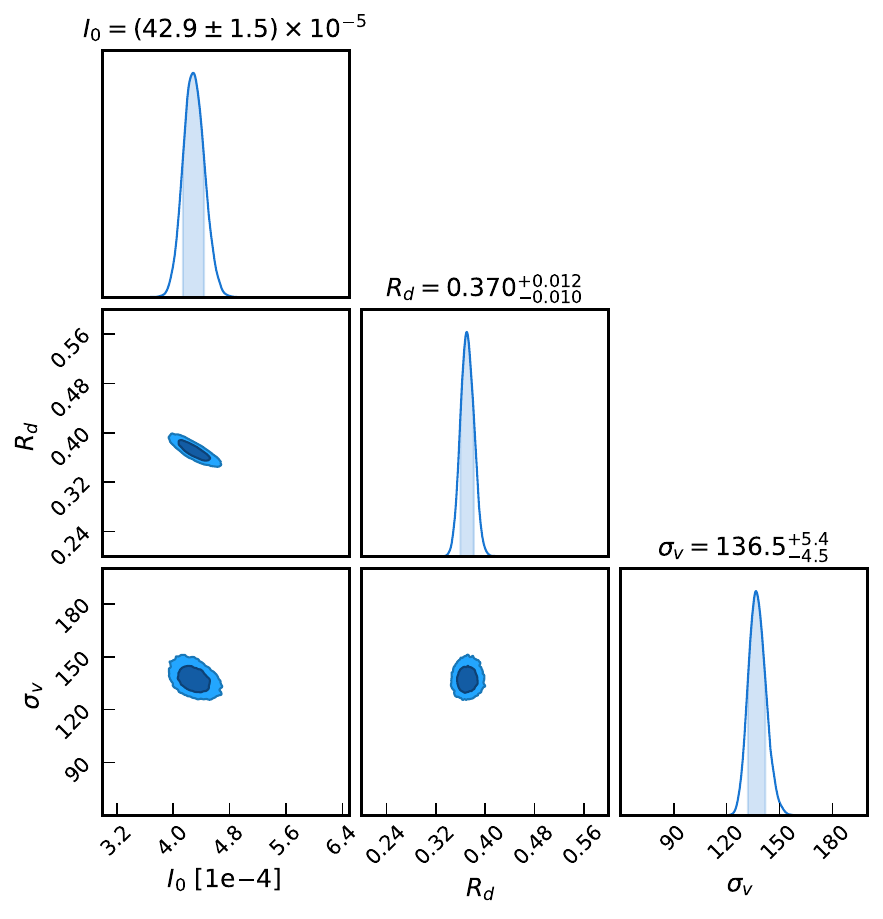}
    \caption{Posterior distribution for the dispersion--dominated model without SMBH, where $I_0$ is the central [\ion{C}{2}] emission in $Jy\, \rm{beam}^{-1}$, $R_d$ is the scale length of the exponentially declining [\ion{C}{2}] profile in kpc, and $\sigma_v$ the spatially constant velocity dispersion in $\kms{}$. To make the graph less crowded, we do not show the posterior for the 3D center of the source ($x,y,\Delta v$) which is tightly constrained to the observed center of the [\ion{C}{2}] emission. }
    \label{fig:posterior_Bulge_noBH}
\end{figure*}

\begin{figure*}
    \centering
    \includegraphics[width=0.85\textwidth]{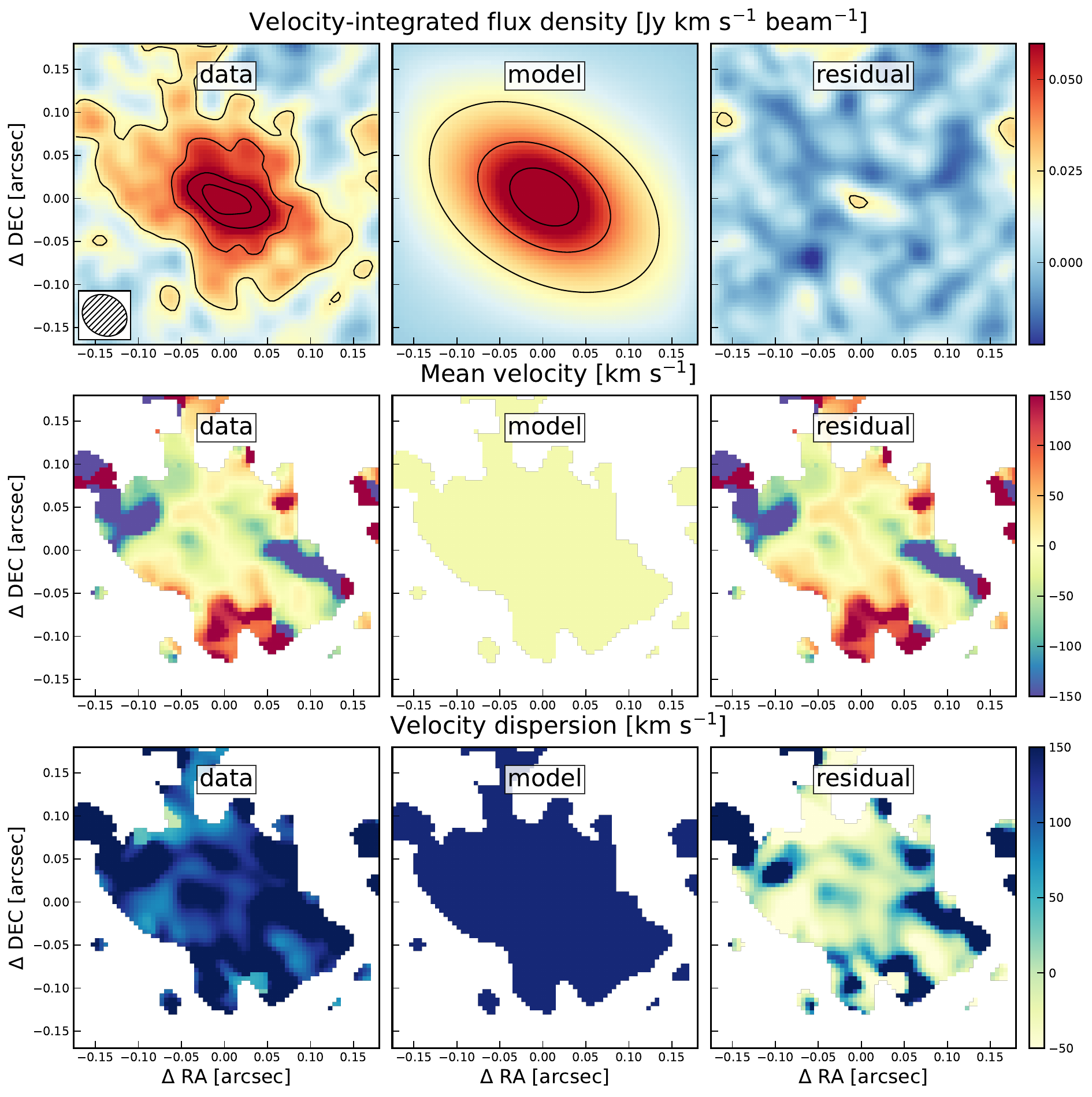}
    \caption{Same as Fig.~\ref{fig:Bulge_noBH} but for the best-fit thin disk model.}
    \label{fig:ThinDisk}
\end{figure*}

\begin{figure*}
    \centering
    \includegraphics[width=0.9\textwidth]{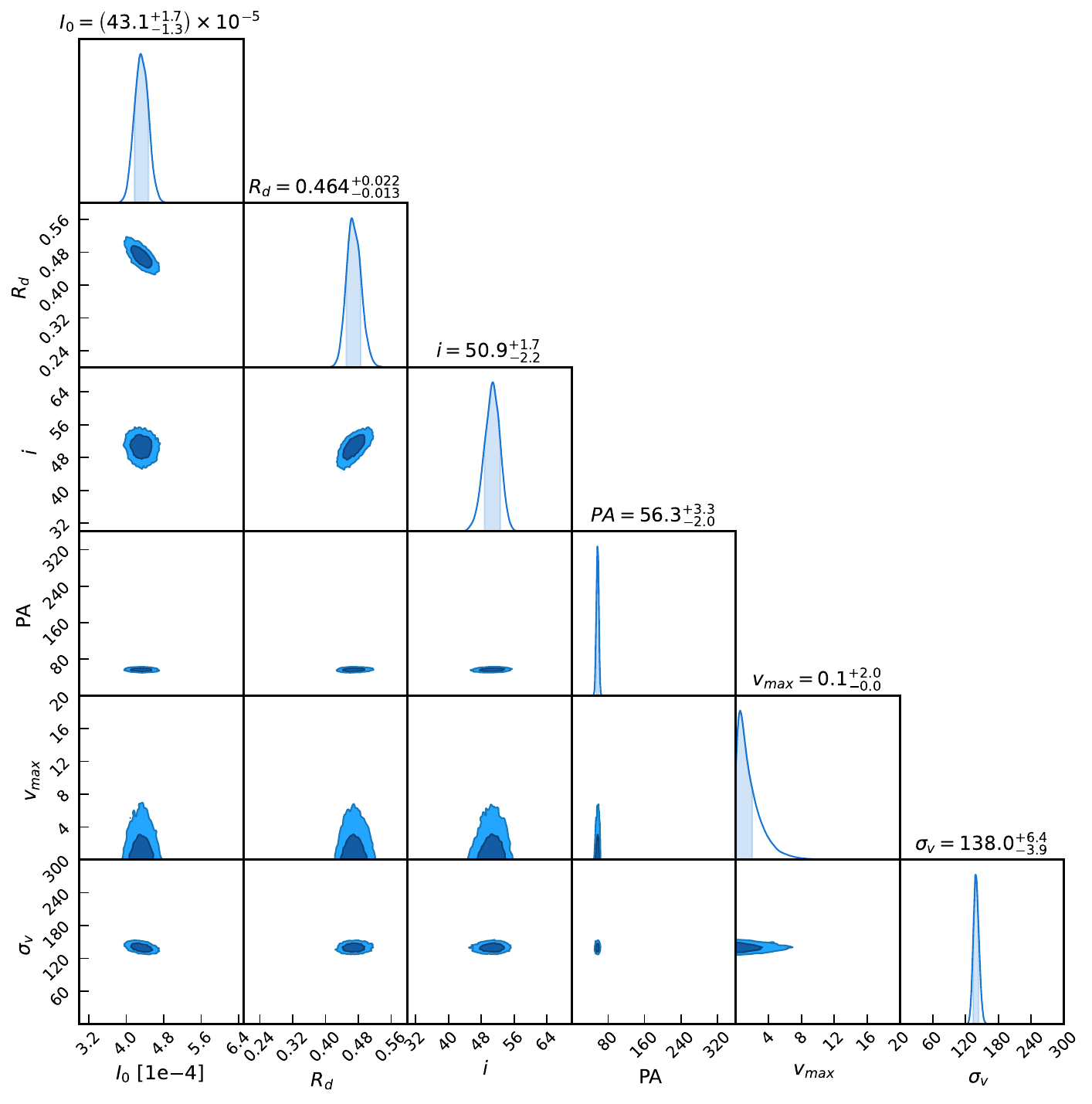}
    \caption{Same as Fig.~\ref{fig:posterior_Bulge_noBH}, but for a thin disk model. The maximum rotation velocity ($v_{\rm{max}}$, in $\kms{}$) is constrained to be $< 4.3\ \kms{}$ at the $2\sigma$ level.}
    \label{fig:posterior_ThinDisk}
\end{figure*}

We here present the four main models discussed in the main text, as well as their residuals and full parameter posterior distributions. The best-fit dispersion--dominated model without a SMBH is shown in Fig.~\ref{fig:Bulge_noBH} and the corresponding posterior distribution in Fig.~\ref{fig:posterior_Bulge_noBH}. The best-fit thin disk model, with a fiducial PA and inclination derived from the major and minor axis of the [\ion{C}{2}] emission, is shown in Fig.~\ref{fig:ThinDisk} and the parameter posterior distribution posterior distribution in Fig.~\ref{fig:posterior_ThinDisk}.

\begin{figure*}
    \centering
    \includegraphics[width=0.85\textwidth]{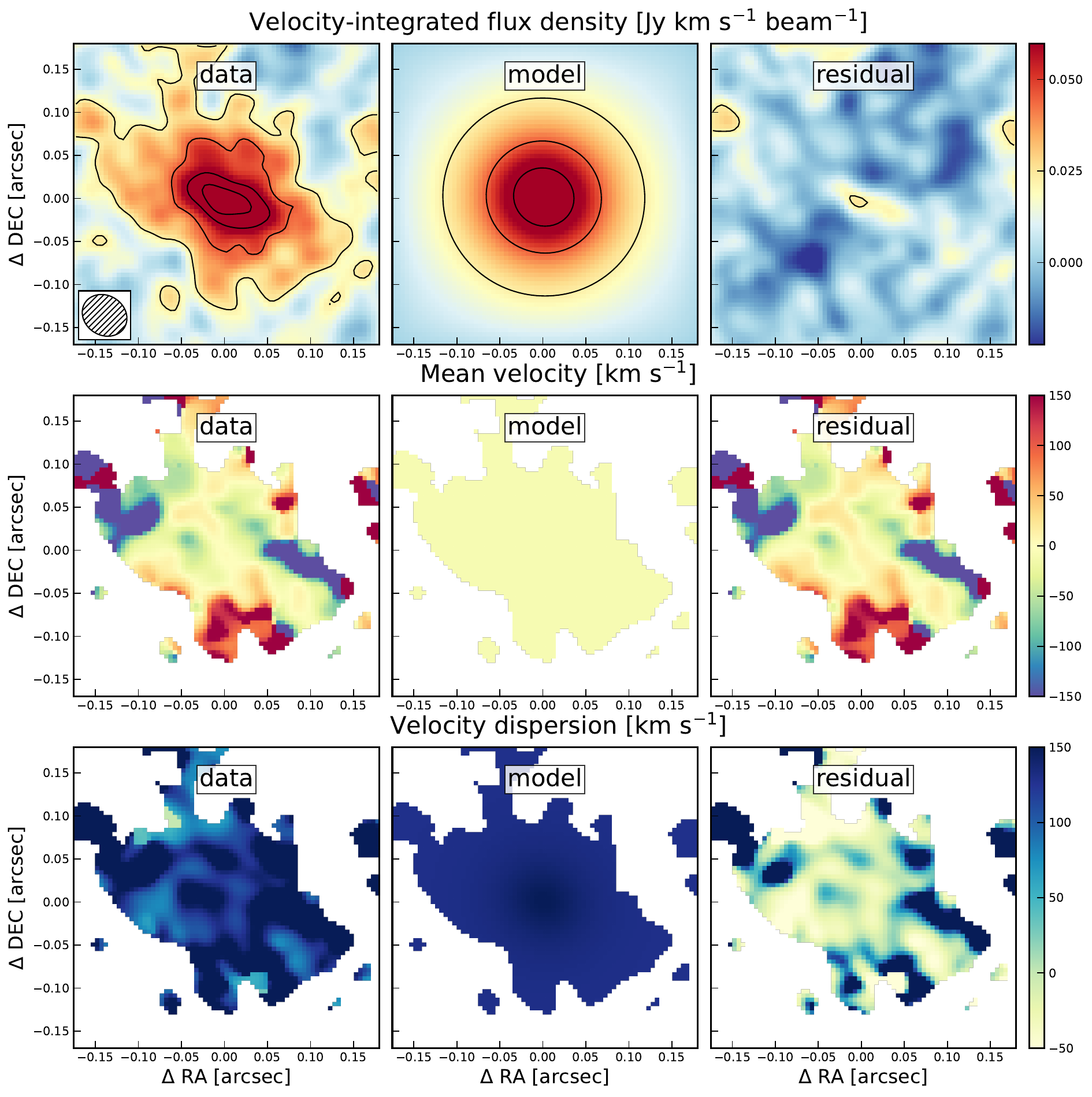}
    \caption{Same as Fig.~\ref{fig:Bulge_noBH} but for dispersion--dominated+SMBH model.}
    \label{fig:Bulge_plusBH}
\end{figure*}

\begin{figure*}
    \centering
    \includegraphics[width=0.7\textwidth]{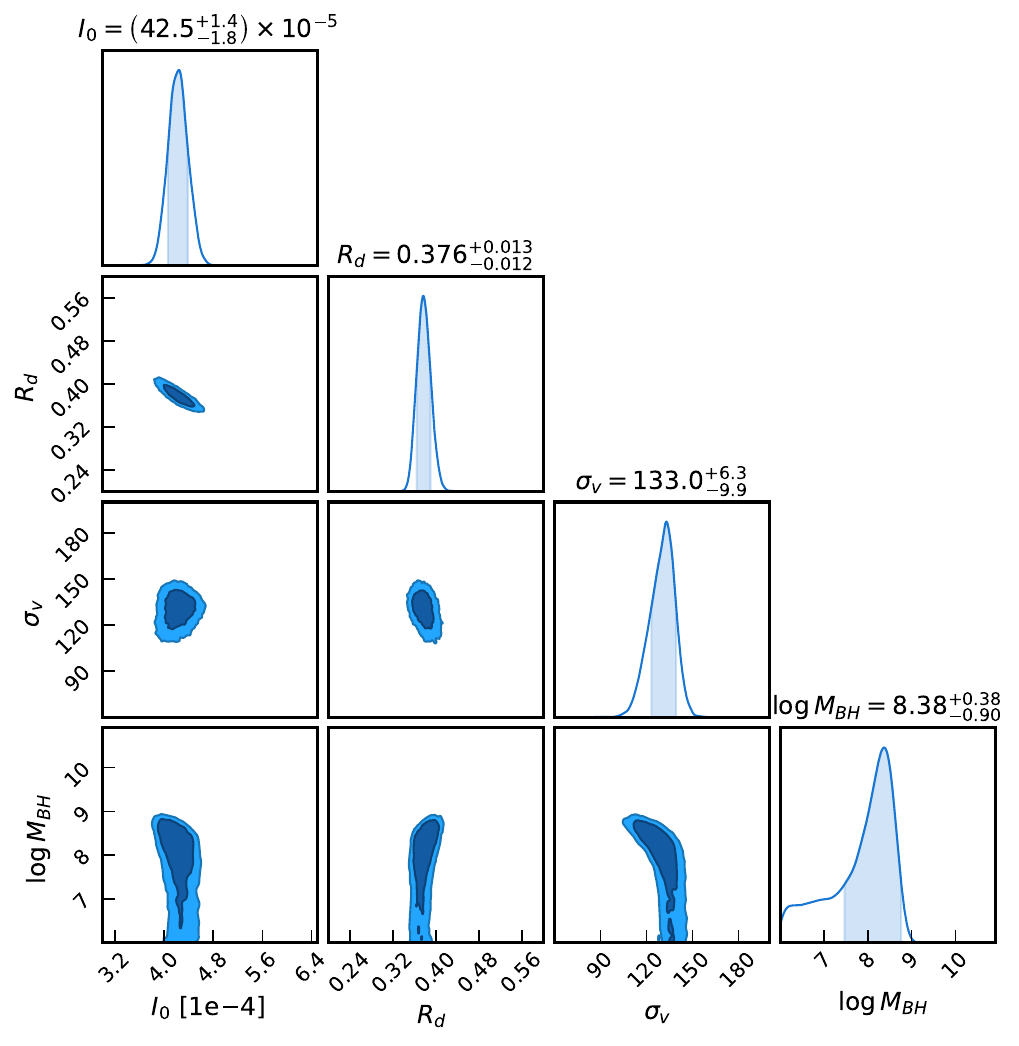}
    \caption{Same as Fig.~\ref{fig:posterior_Bulge_noBH}, but for the dispersion--dominated model including a SMBH.  }
    \label{fig:posterior_Bulge_plusBH}
\end{figure*}

\begin{figure*}
     \centering
     \includegraphics[width=0.9\textwidth]{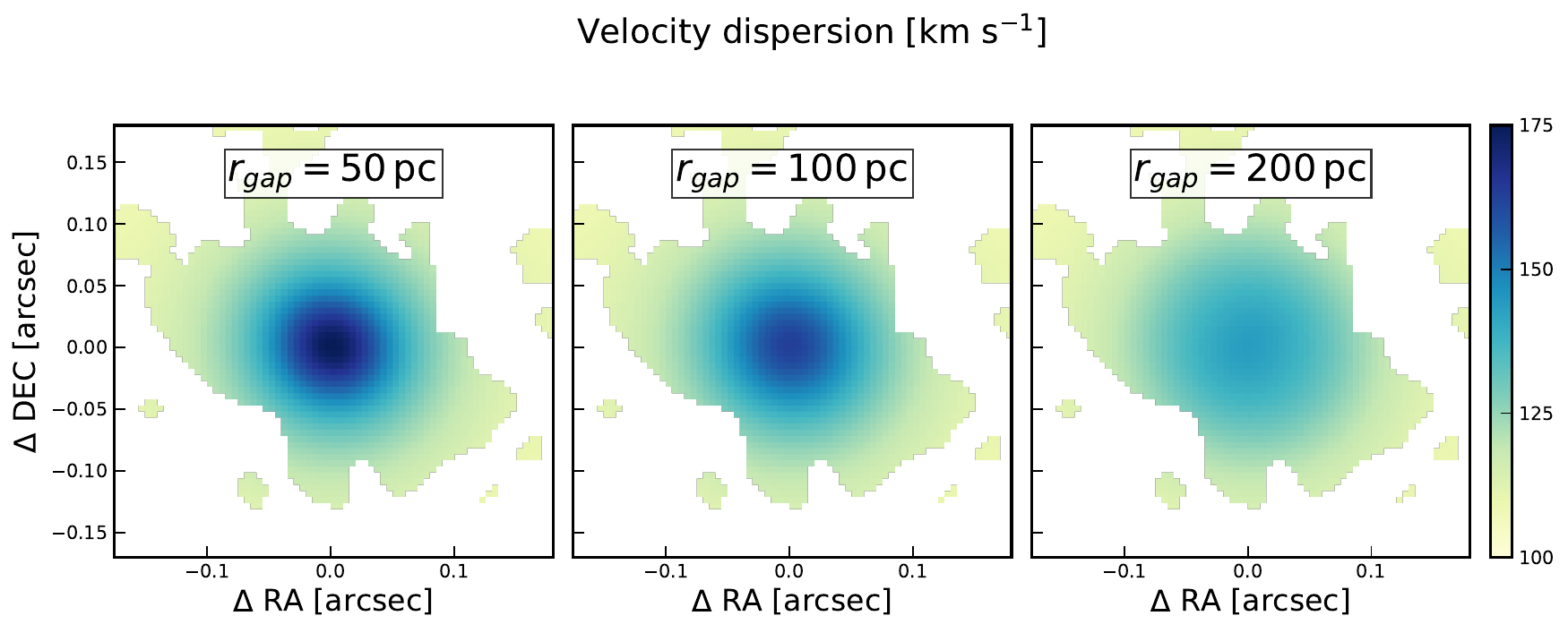}
     \caption{Velocity dispersion of the [\ion{C}{2}] line for different ‘dispersion--dominated+SMBH+gap' model for various central gap sizes. The model is constructed as a dispersion--dominated source with a central $10^{9}M_\odot$ SMBH, and a central gap in the intensity profile ($r_{\rm{gas}}$). The central increase in the velocity dispersion due to the SMBH can be decreased if the gap is large $>200\ \rm{pc}$, but such configurations are excluded as our observations resolve these scales and the [\ion{C}{2}] emission does not present a central dip. }
     \label{fig:dispersion_gaps_models}
 \end{figure*}

\begin{figure*}
    \centering
    \includegraphics[width=0.85\textwidth]{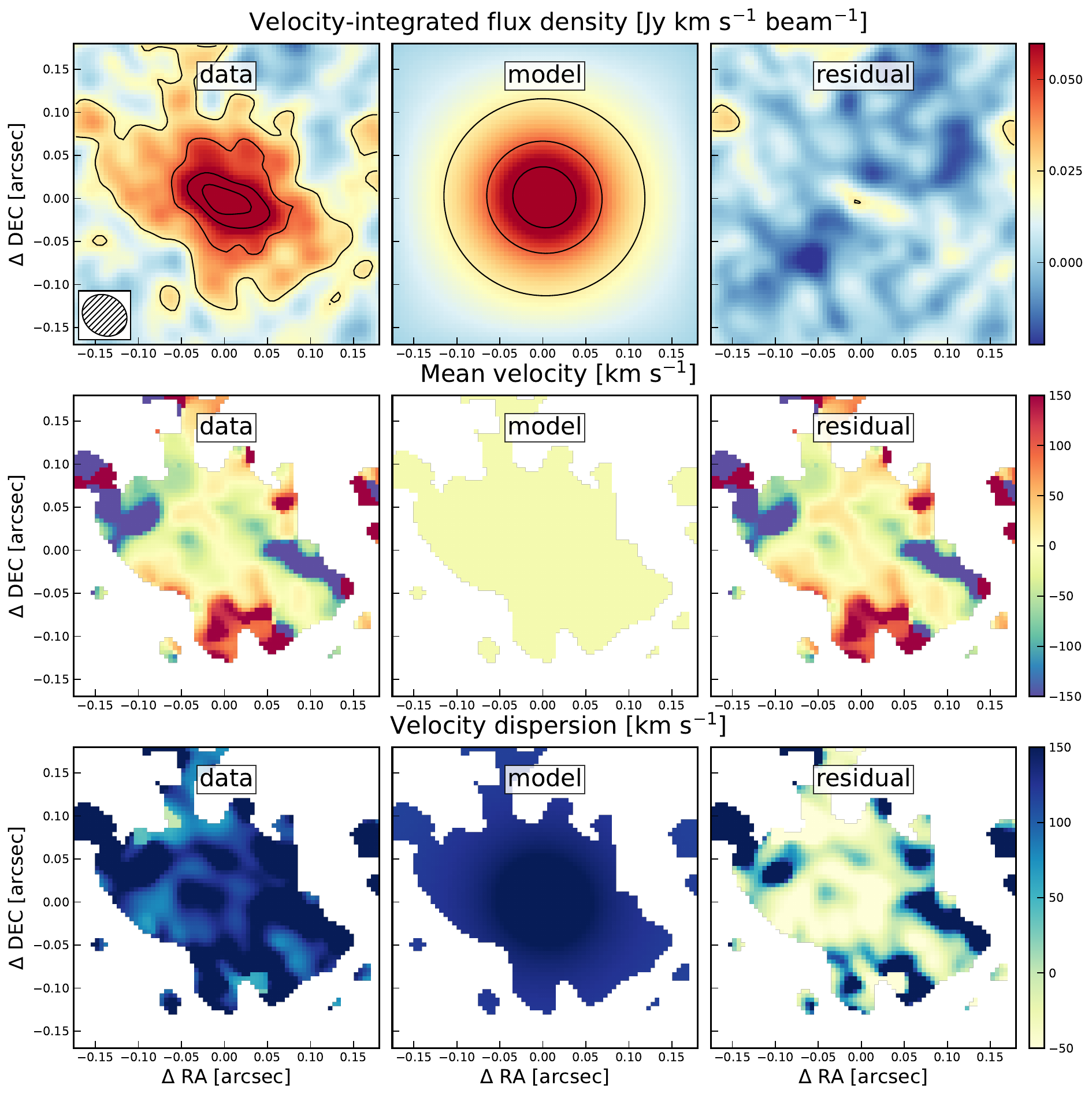}
    \caption{Same as Fig.~\ref{fig:Bulge_noBH} but for the dispersion--dominated+SMBH+gap model. This model has an exponentially declining intensity profile with a central gap and fixed black hole mass of $1.1\times10^{9}\ M_\odot$.}
    \label{fig:Bulge_plusBH_Gap}
\end{figure*}

\begin{figure*}
    \centering
    \includegraphics[width=0.7\textwidth]{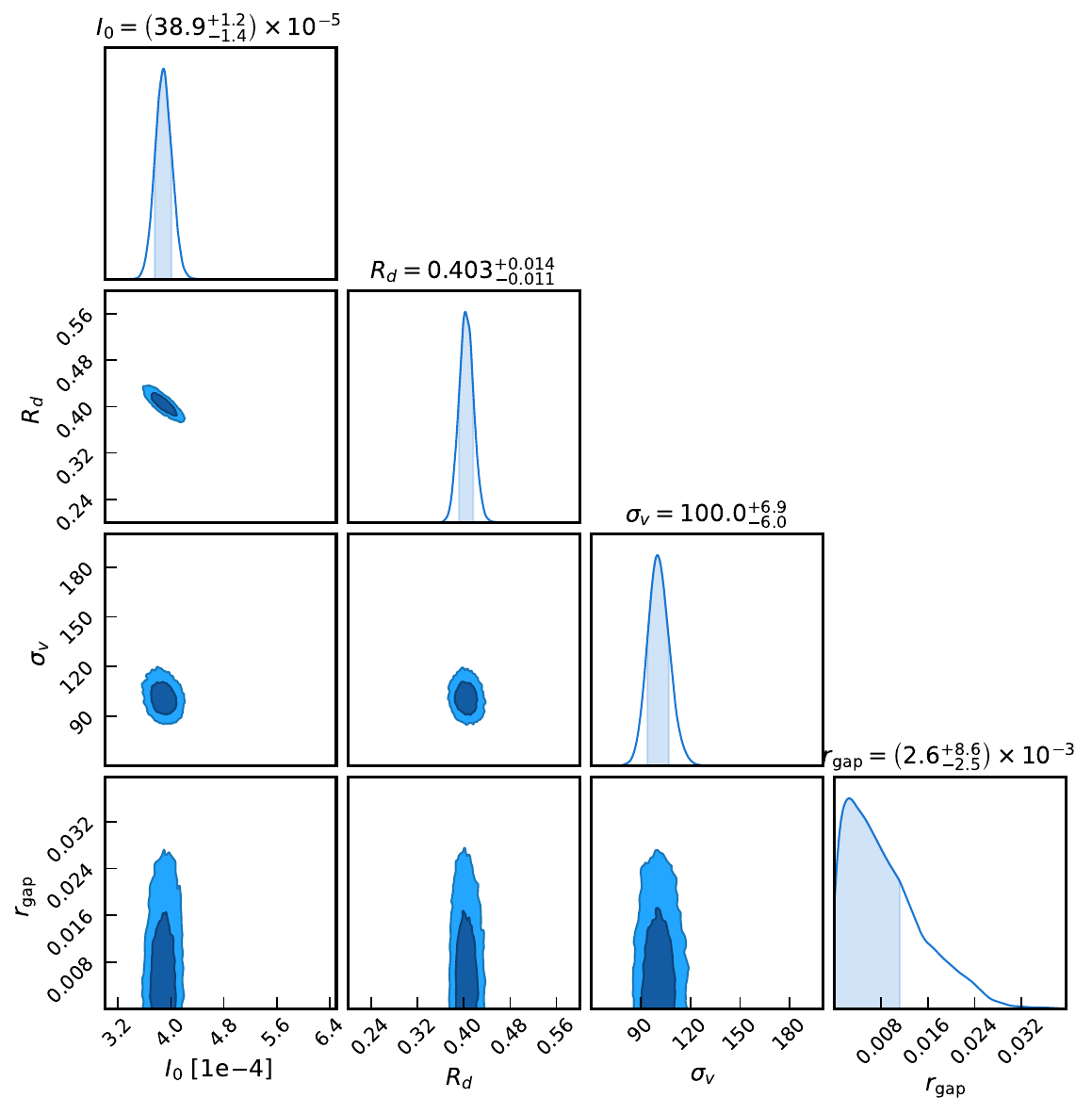}
    \caption{Same as Fig.~\ref{fig:posterior_Bulge_noBH}, but for the dispersion--dominated model including a SMBH with mass $1.1\times10^{9}\ M_\odot$ (fixed) and a central gap (given in kpc)  in the [\ion{C}{2}] emission profile.}
    \label{fig:posterior_Bulge_plusBH_Gap}
\end{figure*}

The bulge dispersion model with a SMBH is shown in Fig.~\ref{fig:Bulge_plusBH} and \ref{fig:posterior_Bulge_plusBH}. We show different realisations of velocity dispersion in the model with a central gap in the intensity profile, as well as the best-fit model and full posterior in Fig.~\ref{fig:dispersion_gaps_models}, \ref{fig:Bulge_plusBH_Gap} and \ref{fig:posterior_Bulge_plusBH_Gap}, respectively. 

\end{document}